\documentstyle[preprint,aps]{revtex}
\begin{document}
\preprint{OITS-549}
\draft
\title{A LIGHT $Z'$ BOSON\footnote
{Work supported in part by the Department of Energy Grant No.
DE-FG06-85ER4022}}
\author{XIAO-GANG HE\footnote{ Talk presented at the Eighth Meeting of the
American Physical Society,
Division of Particles and Fields (DPF'94), Albuqurque, New Mexico, August 2-6,
1994.}}
\address{ Institute of Theoretical Science\\
 University of Oregon\\ Eugene, OR 97403, USA}
\date{August, 1994}
\maketitle
\begin{abstract}
A model of a light $Z'$ boson based on gauged
$L_\mu - L_\tau$ $U(1)$ symmetry is constructed. The $Z'$ boson mass
is constrained to be in the range of 0.8 to 1 GeV from
Z and $Z'$ mass relation, g-2 of muon, and tau decays. The two body decay
$\tau \rightarrow \mu Z'$ is possible. This will provide a
striking signature to test the model.
\end{abstract}
\newpage

The Minimal Standard $SU(3)_C\times SU(2)_L\times U(1)_Y$ Model (MSM)
contains several
global symmetires: the baryon number, and generational lepton number
symmetries. The simplest way to have a new neutral gauge bosn is to gauge
some of these global symmetries. From anomaly cancellation requirement, the
only gaugeable symmetries are:
i) $L_e - L_\mu$, ii) $L_e-L_\tau$, and iii) $L_\mu - L_\tau$, if the
fermion spectrum is not enlarged.
The gauge symmetry breaking scale may or may not be related to the
electroweak symmetry breaking. In the case where the breaking scale is not
related to the
electroweak symmetry breaking, $Z'$ can be made arbitrarily heavy
and therefore the effect at low energy can satisfy experimental
constraints\cite{Le-Lmu}.
 In the case the symmetry breaking scale is related
to the electroweak scale,  $Z'$ can not be arbitrarily
heavy. If $Z'$ couples to the first generation, there
are very stringent bounds on the couplings and may already be ruled out.
For the model in which $Z'$ only couples to $\mu$ and $\tau$,
the constraints from
low energy physics may be very weak. Further more if $Z'$ does not mix with Z,
the precise measurements for Z related observables will not constrain the
model.
Experiments may have missed the chance to detect such $Z'$ boson. In the
following, we construct such a model and study the experimental
constraints\cite{he}.

We consider a model based on $SU(3)_c\times SU(2)_L\times U(1)_Y\times
U(1)_{L_\mu - L_\tau}$ with the same fermion particle contents in the MSM.
All quarks and the first generation of leptons do not transform under the new
$U(1)$ gauge symmetry. The second and third generation
leptons have the following quantum numbers
\begin{eqnarray}
& \ell_{2L} \sim (1,2)(-1,2a),\qquad e_{2R} \sim
(1,1)(-2,2a),&\ \nonumber\\
& \ell_{3L} \sim (1,2)(-1,-2a),\qquad e_{3R} \sim
(1,1)(-2,-2a).
\end{eqnarray}
To ensure that there is no Z-$Z'$ mixing, we impose a discrete symmetry under
which
\begin{equation}
\ell_{2L} \leftrightarrow \ell_{3L},\quad
e_{2R} \leftrightarrow e_{3R},\quad B^{\mu} \leftrightarrow B^{\mu}
\quad {\rm and}\quad Z'^{\mu}
\leftrightarrow -Z'^{\mu},
\end{equation}
where $B^{\mu}$ and $Z'^{\mu}$ are
the gauge fields for U(1)$_Y$ and U(1)$'$, respectively.
This discrete symmetry will also eliminate one free parameter from the kinectic
mixing of the two U(1) gauge fields\cite{foot-he}.

The Yukawa coupling Lagrangian is
\begin{equation}
{\cal L}_{\rm Yuk} = \lambda (\overline{\ell}_{2L} e_{2R} \phi_1 +
\overline{\ell}_{3L} e_{3R} \phi_1)\
 + \lambda' (\overline{\ell}_{2L} e_{3R} \phi_2 +
\overline{\ell}_{3L} e_{2R} \phi_3) + {\rm H.c.},
\label{yuk}
\end{equation}
where the Higgs doublet transform as,
\begin{eqnarray}
\phi_1 \sim (1,2)(1,0),\;\phi_2 \sim (1,2)(1,4a),\;
 \phi_3 \sim (1,2)(1,-4a).
\end{eqnarray}
Under the discrete symmetry $\phi_1 \leftrightarrow \phi_1$ and $\phi_2
\leftrightarrow \phi_3$.

 We adopt, to begin with,
that range of parameters in the Higgs potential which maintains the discrete
symmetry as
exact. This has been shown to be possible\cite{he}. This will guarantee no
Z-$Z'$ mixing to all orders. The
vacuum expectation value pattern required is
\begin{equation}
\langle\phi_1\rangle \equiv u_1\ (\neq 0\ {\rm in\ general})\quad
{\rm and}\quad |\langle\phi_2\rangle| = |\langle\phi_3\rangle| =
u_2 \neq 0.
\end{equation}
The $Z$ and $Z'$ masses are then given by,
\begin{equation}
m^2_Z = {1 \over 2} (g_1^2 + g_2^2) (u_1^2 + 2 u_2^2)\quad {\rm
and}\quad m^2_{Z'} = 16 a^2 s^2_W (g_1^2 + g_2^2) u_2^2.
\label{mass1}
\end{equation}

In the mass eigenstate basis, the $Z'$-lepton couplings are given by
\begin{equation}
{\cal L}^{\ell}_{\rm int} = {{ea} \over c_W}
(\overline{\mu} \gamma^{\mu} \tau
+ \overline{\tau} \gamma^{\mu} \mu) Z'_{\mu}\nonumber\\
+{{ea} \over 2 c_W}
(\overline{\nu}_{\mu} \gamma^{\mu} (1-\gamma_5)\nu_{\tau}
+ \overline{\nu}_{\tau} \gamma^{\mu} (1-\gamma_5)\nu_{\mu}) Z'_{\mu}.
\end{equation}
It is interesting to note that the couplings are always off-diagonal.

Let us now look at the phenomenology of the model. Among many constraints, we
find the following three very interesting: i) Experimental data on $a_\mu=g-2$
 of muon;
2) $Z'$ and Z mass relation; And iii) Experimental data on tau decays.

At the one loop level, the $Z'$ contribution to $a_{\mu}$ is given by,
\begin{equation}
\Delta a^{Z'}_{\mu} = {\alpha_{\rm em} \over 2\pi} {|a|^2 \over
c^2_W} \left\{ \gamma + 2(\beta - {2B\over C}\gamma)
+ 2M\ln\left({m_\tau \over m_{Z'}}\right) + \delta \right\},
\label{g-2Z'}
\end{equation}
where
\begin{equation}
\delta = \left\{
\begin{array}{ll}
{{NC-MB} \over \sqrt{B^2-AC}} \ln\left|
{{A+B+\sqrt{B^2-AC}} \over {A+B-\sqrt{B^2-AC}}}\right|
& \mbox{if $B^2>AC$;} \\
2{{NC-MB} \over \sqrt{AC-B^2}} \tan^{-1}\left[{\sqrt{AC-B^2}
\over {A+B}}\right] & \mbox{if $B^2<AC$.}
\end{array} \right.
\end{equation}
In these equations, $\alpha_{\rm em}$ is the fine-structure constant,
\begin{eqnarray}
&\alpha \equiv 2({m_{\tau} - m_{\mu}})/m_{\mu}, &\ \nonumber\\
&\beta \equiv 3 - 2(m_{\tau}/ m_\mu) +
{1\over 2 }\left((m_{\tau}/ m_{\mu})+1\right)
(m_{\tau}-m_{\mu})^2/ m_{Z'}^2,&\ \nonumber\\
&\gamma \equiv -1 - {1\over 2}(m_{\tau}-m_{\mu})^2/m^2_{Z'},&\ \nonumber\\
&A \equiv m^2_{Z'},\quad B \equiv (m^2_{\tau}-m^2_{\mu}-m^2_{Z'} )/2,
\quad C \equiv m^2_{\mu},&\ \nonumber\\
&M \equiv \alpha - {2B\over C}(\beta-{2B\over C}\gamma)\quad {\rm and}
\quad N \equiv -{A\over C}(\beta-{2B\over C}\gamma).&\
\end{eqnarray}
 In the $m_{Z'} \gg m_{\tau}$ limit, Eq.~(\ref{g-2Z'})
reduces to the simple result that
\begin{equation}
\Delta a^{Z'}_{\mu} \simeq {\alpha_{\rm em} \over 2\pi} {|a|^2 \over
c^2_W}{{2m_{\mu}m_{\tau}} \over m^2_{Z'}} = {m_{\mu}m_{\tau} \over
64\pi^2u^2_2}
\end{equation}
which is independent of $|a|$ and at best about an order of magnitude
too large given that $u_2$ is constrained by the electroweak scale. A detailed
study find that if $m_{Z'}$ is a few GeV or less, for small $|a|$ (less than
$10^{-2}$),
it is possible to satisfy the bound\cite{pdg} on $\Delta a_\mu < 10^{-8}$.

We next consider the constraint coming from the gauge boson masses.
By using the expressions for the masses of Z and $Z'$, and require $u_1^2 > 0$,
we find
\begin{equation}
|a| > {1\over 4\sin\theta_W}{m_{Z'}\over m_Z} \simeq \left({m_{Z'}
\over 175.33 \ \hbox{GeV}}\right).
\label{bd}
\end{equation}
For values of $a$ and $m_{Z'}$ close to saturating the bound, $u_1$ becomes
small so that it is possible for certain quark Yukawa couplings to become
nonperturbative.

The above constraint requires that for a given $Z'$ mass, the parameter $|a|$
has to be greater than certain value.
When the above two constraints are combined there is a small overlap
region allowed (roughly $m_{Z'} < 2.5$ GeV).

Let us first consider the case where $\tau \to \mu Z'$ is not allowed.
Although this dramatic two-body decay does not occur, the off-shell $Z'$
contributes to $\tau^- \to \mu^- \overline{\nu}_{\mu}
\nu_{\tau}$. For this mode, the $Z'$
contribution coherently adds with the standard $W$-boson contribution
yielding
\begin{eqnarray}
R & \equiv & {\Gamma(\tau^- \to \mu^- \overline{\nu}_{\mu} \nu_{\tau}) \over
\Gamma(\tau^- \to \mu^- \overline{\nu}_{\mu} \nu_{\tau})_{\rm SM}}
\nonumber \\
& = & 1 - \xi \left[ 2k(k+1) - {5 \over 6}
- k^2(2k+3)\ln \left|{1+k \over k} \right| \ \right]\nonumber\\
& + & { 1\over 4} \xi^2 \left[ 2(2k+1) + k{2k+3 \over k+1} - 6k(k+1) \ln
\left|{1+k \over k}\right| \ \right],
\label{3body}
\end{eqnarray}
where
\begin{equation}
k \equiv {m^2_{Z'} \over m^2_{\tau}} - 1\quad {\rm and}\quad \xi \equiv
{\sqrt{2} \over G_Fm^2_{\tau}} {4\pi\alpha_{\rm em} \over c^2_W} |a|^2.
\end{equation}
 The experimental constraint \cite{pdg}
\begin{equation}
|R-1| < 0.04
\end{equation}
in fact closes the $(m_{\tau}-m_{\mu}) < m_{Z'} < 2.5$ GeV window. The allowed
windows for $m_{Z'}$ are $ < 0.2$ GeV and $0.8\sim 1$ GeV. There is also a
minute region at $m_{Z'}\sim 1.2$ GeV.

We are left to consider the kinematic region which permits the
two-body decay mode $\tau \to \mu Z'$.
The Mark III and ARGUS collaborations \cite{exp} have set limits on two-body
decay modes for $\tau$. These experimental groups specifically analysed
the process $\tau \to \mu +\ \hbox{Goldstone Boson}$ and found that the
ratio
\begin{equation}
P={\Gamma(\tau \rightarrow \mu Z') \over
\Gamma(\tau \rightarrow \mu\overline{\nu_\mu} \nu_\tau)}
< 0.033, \quad \hbox{for} \ m_{Z'} \leq 0.1 \ \hbox{GeV},
\label{2bb}
\end{equation}
where the Goldstone boson has been replaced by $Z'$.
(Without going into a detailed reanalysis of the experiment,
we expect the above experimental bound to be approximately valid
for our case where the final state boson has spin-1.)
This bound rises up to $0.071$ for $m_{Z'} = 0.5$ GeV. The
ratio P in this model is given by
\begin{equation}
{96\over\sqrt{2}}\pi^2 \tan^2\theta_W
{m_W^2\over G_F m_{\tau}^4} |a|^2\left\{1+{(m_\mu^2-2m_{Z'}^2)\over
m_{\tau}^2}-6{m_\mu\over m_\tau}
+{(m_\mu^2-2m_{Z'}^2)^2\over m_{\tau}^2 m_{Z'}^2}\right\}PS ,
\label{2body}
\end{equation}
where
\begin{equation}
PS=\sqrt{1-{(m_\mu+m_{Z'})^2\over m_{\tau}^2}}
\sqrt{1-{(m_\mu-m_{Z'})^2\over m_{\tau}^2}}.
\end{equation}
Using this expression, we find that $m_{Z'} < 0.5$ GeV is ruled out.

In summary, when all the constraints have been combined,
much of the parameter space is ruled out. The
remaining allowed regions are for $0.8 < m_{Z'} < 1$ GeV ($|a|$ varies
between about $0.004$ and $0.007$) and $m_{Z'}$
around $1.2$ GeV. It should be noted that we have
taken the two-body constraint at face value, i.e., it applies for
values of $m_{Z'}$ up to $0.5$ GeV. Actually, the ARGUS experiment
is supposed to be able to search for the two-body decay mode for values
of $m_{Z'}$ up to about $1.53$ GeV, given the experimental cuts and
efficiencies. If the current trend of the two-body constraint continues
beyond $0.5$ GeV (the precise bound will obviously vary with the mass of
$Z'$ and becomes several orders of magnitude less severe near threshold)
then the remaining allowed windows will be closed and the model will be
ruled out.

We have check the case when the discrete symmetry is spontaneously broken. The
constraints are less stringent, but not enough to
dramatically change anything significantly.

I thank Drs. Foot, Lew and Volkas for collaboration on this subject.

\end{document}